\documentclass[twocolumn, twocolappendix]{aastex631}
\usepackage[T1]{fontenc}
\usepackage{amsmath}
\usepackage{wrapfig}
\usepackage{svg}
\usepackage{amssymb}
\usepackage{graphicx}
\usepackage{subfigure}
\usepackage{xcolor}
\usepackage{float}
\usepackage{epstopdf}
\usepackage{txfonts}
\usepackage{mathrsfs}
\usepackage{graphics}
\usepackage{keyval}
\usepackage{url}
\usepackage{multirow}
\usepackage{bm}
\usepackage[figuresright]{rotating}

\newcommand{\Rmnum}[1]{\expandafter\@slowromancap\romannumeral #1@}
\allowdisplaybreaks[3]

\newcommand \beq{\begin{equation}}
\newcommand \eeq{\end{equation}}
\newcommand \bey{\begin{eqnarray}}
\newcommand \eey{\end{eqnarray}}

\newcommand \kpc{\, {\rm kpc} }
\newcommand \Mpc{\, {\rm Mpc} }
\newcommand \msun{{\rm M}_\odot}
\newcommand \Msun{{\rm M}_\odot}

\newcommand \kms{\, {\rm km \, s}^{-1} }

\newcommand \conveff{{f_\star}}
\newcommand \nden{{\mathfrak n}}

\begin{document}

\title{Measuring Mass of Gas in Central Galaxies using weak lensing and satellite kinematics in MOND}

\author[0009-0005-7751-7280]{Li~Ma}
\affiliation{Department of Astronomy, University of Science and Technology of China, Hefei 230026, China}
\affiliation{School of Astronomy and Space Science, University of Science and Technology of China, Hefei 230026, China}
\author[0000-0002-9272-5978]{Ziwen~Zhang}
\affiliation{Department of Astronomy, University of Science and Technology of China, Hefei 230026, China}
\affiliation{School of Astronomy and Space Science, University of Science and Technology of China, Hefei 230026, China}
\author[0000-0002-4911-6990]{Huiyuan~Wang}
\affiliation{Department of Astronomy, University of Science and Technology of China, Hefei 230026, China}
\affiliation{School of Astronomy and Space Science, University of Science and Technology of China, Hefei 230026, China}
\author[0000-0002-1378-8082]{Xufen Wu}\altaffiliation{Corresponding author: xufenwu@ustc.edu.cn}
\affiliation{Department of Astronomy, University of Science and Technology of China, Hefei 230026, China}
\affiliation{School of Astronomy and Space Science, University of Science and Technology of China, Hefei 230026, China}

\begin{abstract}
In the Milgromian dynamics (MOND) framework, the dynamical mass of a galaxy is fully determined by its baryonic matter distribution. Using this framework, we fit the distribution of cold and hot gas halos$-$ focusing on the hot gas$-$ around SDSS central galaxies, utilizing weak lensing signals from the DECaLS survey. The central galaxies are classified into total sample and star-forming sample. Hot gas halo densities nearly follow Plummer's profile for both samples across all mass bins out to extended radii. We then demonstrate the rotation curves of the galaxy samples. Furthermore, the stellar fraction, $M_*/(M_* + M_{\rm g})$, is between 0.3 and 0.8 in all mass bins of the star-forming sample, which is higher than in the total sample. Additionally, we use the satellite kinematics method in MOND to verify our findings from the weak lensing method. We find good agreement between the two methods, indicating that weak lensing signals reliably measure the dynamical mass of central galaxies and can constrain the distribution of missing baryons in galaxy clusters. Combining both methods, we discover a baryonic mass to line-of-sight velocity dispersion of satellites ($M_{\rm b}$-$\sigma_{\rm s}$) relation. More sophisticated models, such as Osipkov-Merritt anisotropy profiles, were found unnecessary, as simple isotropic or mildly radially anisotropic MOND models align well with the observed $M_{\rm b}$-$\sigma_{\rm s}$ relation. Moreover, the isotropic model remains consistent with this relation even when considering external fields from large-scale structures.
\end{abstract}



\section{Introduction}
The multiphase gaseous components in galaxies are ideal tracers of {their} gravitational potential. The measurements of the emission lines from the cold gas, including H$\alpha$, HI, CO and maser emissions, are the most important methods to study the rotation curves of spiral galaxies {\citep[see the reviews by][]{Sofue_Rubin2001,Sofue2017,Faber_Gallagher}}. 
The cold molecular and atomic gas contributes a large portion to the circular velocities of galaxies. In the central few ${\rm kpc}$ of spiral galaxies, over $90\%$ of the cold gas is molecular gas \citep{Sofue+1995}. In contrast, at larger radii, the cold gas is predominantly HI gas \citep{Bosma1981}. Observations of HI gas have shown that rotation curves remain flat at these larger radii, extending to a few times beyond the optical disks \citep[e.g.,][]{Bosma1981,Begeman1989}. {Thus,} the distribution of HI gas in galaxies significantly influences the rotation curves at large radii. There are many efforts on mapping the radial distribution of HI gas in galaxies \citep[e.g.,][]{vanderhulstWesterborkHISurvey2001,swatersWesterborkHISurvey2002}. In these works, a homogeneous radial profile has been found in the outer regions of dwarf and spiral galaxies \citep{wangObservationalTheoreticalView2014,wangNewLessonsSize2016}. 

In addition to cold baryons such as stars, cold gas, and black holes within galaxies, observations indicate that the primary baryonic component in the low {redshift} universe is the hot intergalactic medium \citep{Shull+2012}. 
Numerous groups have studied the radial distribution of X-ray emitting hot gas around the Milky Way and other external galaxies \citep{vikhlininChandraSampleNearby2006,millerSTRUCTUREMILKYWAY2013,bogdanProbingHotXRay2017}. The X-ray profiles of star-forming and quiescent galaxies have been measured out to a few hundred $\rm kpc$, revealing distinct differences in their radial projected luminosity profiles \citep{comparatEROSITAFinalEquatorial2022}. These X-ray profiles are important for studying galaxy mass {distributions} and may explain a significant portion of the missing baryons \citep{bregmanExtendedDistributionBaryons2018a,bogdanXRayHalosMassive2022,nicastroXRayDetectionGalaxy2023}.

In the framework of $\Lambda$CDM, galaxies are hosted by dark matter halos. Galaxy-galaxy lensing (hereafter GGL) provides {a} direct measurement of the total mass distribution of galaxies{, including their} dark matter halos \citep{yangWeakLensingGalaxies2006,luoGalaxyGalaxyWeaklensing2018,Zhang+2022}. Recently, GGL signals have been utilized to derive the radial acceleration relation (RAR) of galaxies \citep{brouwerWeakLensingRadial2021,Mistele+2024}. Alternatively, the kinematics of satellite galaxies {(SK)} can also be used to probe the total mass distribution of central galaxies \citep[e.g.,][]{vandenBosch+2004,More+2009a,More+2009b,More+2011,Pei_Wang2018,Lange+2019,Li+2020,Zhang+2024b}.
\cite{Zhang+2022} has introduced a method combining GGL and SK to investigate the gas-to-stars conversion efficiency in central galaxies. Scaling relationships between halo masses and galaxy properties including stellar masses have been further studied using the same method \citep{Zhang+2024b}. 

On the other hand, as an alternative to the $\Lambda$CDM models, Milgrom's MOdified Newtonian Dynamics (MOND; \citealp{Milgrom1983a}) is proposed to solve the mass discrepancy problem without dark matter.
The MOND theory has been tested on various scales ranging from a few parsecs to a few Gpc, and is found to be favored by many observations (\citealp[e.g.,][]{sandersModifiedNewtonianDynamics2002,famaeyModifiedNewtonianDynamics2012}, {and also see the review of \citealt{banikGalacticBarsHubble2022}, hereafter BZ22)}. 
The LoS velocity dispersion profiles for globular clusters in the Milky Way are flat \citep{Scarpa+2007,Lane+2009,Hernandez_Jimenez2012}, which has been predicted in MOND \citep{Milgrom1994}. MOND offers influential alternative interpretations of the kinematics and dynamics of galaxies \citep[e.g.,][]{Milgrom2009,Kroupa+2012,Kroupa2015}. 
On the scale of the cluster of galaxies, MOND faces several challenges. The lensing data requires supplementary dark matter even in the context of modified gravity, for clusters including the Bullet Cluster {1E0657-56} and the dark matter ringed cluster Cl0024+17 {\citep{Clowe+2006,Angus+2007,Jee+2012}. Postulating additional central dark mass in galaxy clusters to salvage MOND predictions creates a tension between insufficient central gravity and excessive outskirts gravity \citep{Li+2023,Kelleher_Lelli2024}. Their analysis reveals that excluding dynamically perturbed clusters A644 and A2319, the RAR for three virialized clusters requires a fixed enhancement of approximately $ 2\pi$ of Newtonian baryonic gravity in low-acceleration regimes. Crucially, centrally concentrated dark matter implementations would imply an inverse-square gravitational law below accelerations in the MOND framework. Moreover, this $2\pi$ factor aligns remarkably with the cosmic baryon-to-total-matter density ratio implied by the cosmic microwave background (CMB) anisotropies in the $\Lambda$CDM framework \citep{Li+2023,Kelleher_Lelli2024}.} In the original relativistic MOND framework, the third peak of the angular power spectrum of the CMB radiation was a significant hurdle \citep{Skordis+2006}. However, the new relativistic MOND theory \citep{Skordis_Zlosnik2021,Skordis_Zlosnik2022} has successfully matched the CMB temperature and polarization angular power spectra. 

{While baryons alone can fit CMB observations, classical MOND frameworks conflict with multiple dynamical constraints from planetary scale to sub-parsec scale: Cassini radiometric data \citep{Desmond+2024}, phase-space distributions of long-period comets and trans-Neptunian objects \citep{Vokrouhlicky+2024}, and wide binary kinematics \citep{Banik+2024}. However, the question of whether MOND effects manifest in wide binary systems remains controversial (\citealp[a review by][]{Hernandez+2024}, and also \citealp{Chae2024,Hernandez_Kroupa2025}).}

In the framework of MOND, galaxies only consist of baryons and the dynamical mass of a galaxy is fully {predictable} once the baryonic distribution is known. Consequently, the galaxy-galaxy weak lensing shear signals directly relate to the baryonic mass distribution in MOND. Weak lensing can be served as a test for MOND {\citep{mortlockEmpiricalConstraintsAlternative2001,mortlockGravitationalLensingModified2001a,milgromTestingMONDParadigm2013,Brouwer+2017,Brouwer+2021}}.
Using the GGL signals, \cite{misteleIndefinitelyFlatCircular2024} found that the rotation curves for a sample of isolated galaxies are flat out to hundreds of kilo-parsecs, which strongly supports MOND. Moreover, MOND has also been tested using satellite kinematics \citep{lokasVelocityDispersionsDwarf2001,angusVelocityDistributionSloan2007,tiretVelocityDispersionEllipticals2007,angusDwarfSpheroidalsMOND2008,klypinTESTINGGRAVITYMOTION2009}. 

Given the direct relation between the dynamical and baryonic mass in MOND, it is possible to predict the distribution of the undetected baryons {from the observed and dynamical mass}. These baryons can be the hot gas around galaxies. Since both the GGL signals and the {motions} of satellite galaxies reflect the dynamical masses of the central galaxies, these two methods can be used to explore the gas distribution within the framework of MOND. In this work, we will use these two methods to model the distribution of gas components in galaxies in MOND, using the observational data from \cite{Zhang+2022}. The baryonic mass model for the galaxies, including stellar, cold, and hot gaseous components, is adapted to fit the GGL data for two samples of galaxies, i.e., the star-forming sample and the total sample containing {both} quiescent and star-forming galaxies. The fraction between the gas and overall baryon mass is calculated for both samples in \S \ref{GGL}. Moreover, the gas mass in both samples is {calculated by subtracting the stellar mass from the dynamical mass (under the assumption that all matter is baryonic), using} the line-of-sight (LoS) velocity dispersion profiles of the satellite galaxies in \S \ref{sk}. A relation between the baryonic mass and the global LoS velocity dispersion of the satellites is studied in \S \ref{msigma}. More sophisticated models are incorporated in \S  \ref{sophisticated}. We summarize our results in \S \ref{conclusions}.

\section{Constraining baryonic mass distribution using GGL}
\label{GGL}
\subsection{Sample selection}
The galaxy samples used in this work \citep{Zhang+2022} are selected from 
the New York University Value Added Galaxy Catalog (NYU-VAGC; \citealp{blantonNewYorkUniversity2005}) of the Sloan Digital Sky Survey Data Release 7 (SDSS DR7; \citealp{abazajianSeventhDataRelease2009}). The halo-based group catalog \citep{yangGalaxyGroupsSDSS2007} identifies the central galaxies. Two samples of central galaxies are selected. One is the total population of 304,162 galaxies{, including both} the quiescent and star-forming galaxies. The other contains 129,278 star-forming galaxies. The Excess Surface Density (ESD) profiles are measured by the Dark Energy Camera Legacy Survey (DECaLS; \citealp{deyOverviewDESILegacy2019}) DR8 imaging data. The stellar masses of the galaxies are obtained from the SDSS photometry data. Each galaxy sample is divided into six sub-samples binned in stellar masses, $M_*$. The binning method segments the logarithm of $M_*$ into equal bins of 0.5 dex. 
The ESD profiles of each sub-sample are obtained by stacking the ESD profiles of individual galaxies within that sub-sample. The mean stellar mass of each sub-sample is then easily calculated.

\subsection{Three-component baryonic mass model}
A baryonic mass model for the central galaxies is required to reproduce the ESD profiles of the galaxy samples in MOND. We present a three-component spherically symmetric model here, including stars, a cold gas halo, and a hot gas halo. 

For the stellar component, we use Hernquist's mass density model as  {follows \citep{hernquistAnalyticalModelSpherical1990}:}
\begin{equation}\label{hernquist}
    \rho_*(r)=\frac{M_*}{2\pi(r/a)(r+a)^3},
\end{equation}
where $M_*$ is the stellar mass of a galaxy {and} $a$ is the scale length which can be estimated from the effective radius of the stellar component by $a\approx R_e/1.8153$. The stellar mass and the effective radius are measured from observations.

For the HI gas component, the surface density profile was suggested by Eq. 9 in \cite{wangObservationalTheoreticalView2014}. Here we derive the three-dimensional mass density, $\rho_\mathrm{cg}(r)$, from the surface density distribution. It follows
\begin{equation}
    \rho_\mathrm{cg}(r)=\frac{\Sigma_0 K_0(r/R_{\rm s})}{\pi R_{\rm s}}.
\end{equation}
In the above equations, $\Sigma_0$ and $R_{\rm s}$ are the central surface density and the scale length of the projected radius for the HI gas, respectively. $K_0$ is the zeroth-order modified Bessel function of the second kind.

The mass density distribution for the hot gas halo, $\rho_\mathrm{hg}(r)$, follows a parametric power law proposed by \cite{cavaliereDistributionHotGas1978}, 
\begin{equation}
    \rho_\mathrm{hg}(r)=\rho_0\left[1+(r/r_{\rm c})^2\right]^{-\frac{3}{2}\gamma},
    \label{model}
\end{equation}
where $\rho_0$ and $r_{\rm c}$ represent the central density and core radius of the hot gas halo, respectively. The parameter $\gamma$ is the power law index for the hot gas halo profile.

The total baryonic mass distribution is the sum of the three components,
\begin{equation}\rho_{\rm b}=\rho_*+\rho_\mathrm{cg}+\rho_\mathrm{hg}.\label{rho_b}\end{equation}
The parameters of both {the} HI gas and the hot gas halo models, including $\Sigma_0,\ R_{\rm s},\ \rho_0,\ r_{\rm c},\ {and}\ \gamma$, will be fitted from the ESD profiles of the galaxy samples in the following \S \ref{ESDfit}.

\subsection{MOND fomulations}
As an alternative to the standard $\Lambda$CDM framework, MOND \citep{Milgrom1983a} predicts precisely the rotation curves of galaxies in the absence of dark matter halos by a modification to gravity, which follows
\beq {\bf g}_N=\mu(X) {\bf g},\eeq 
where ${X\equiv \frac{g}{a_0}}$. When the strength of actual acceleration, $g=|{\bf g}|$, is much greater than Milgrom's constant, ${a_0=1.2\times 10^{-10}~{\rm m}~{\rm s}^{-2}=3700 ~{\rm km}^2 {\rm s}^{-2} \kpc^{-1}}$, the interpolating function $\mu \rightarrow 1.0$. In this limit, the actual acceleration approaches the Newtonian acceleration induced by the baryonic matter. In the weak field limit where $g<< a_0$, $\mu \rightarrow X$. The formulation becomes
\bey\label{eq:aqual}  {\bf g}_N=\mu(X) ({\bf g}-{\bf g}_{\rm ext}), \nonumber \\
X=\frac{|{\bf g}-{\bf g}_{\rm ext}|}{a_0} \eey
when there is an external field, ${\bf g}_{\rm ext}$ {(\citealp{Milgrom1983a}; also see Section~2.4 of BZ22)}.

Several MOND formulations satisfy the conservation laws, such as AUQAL \citep{BM1984} and QUMOND \citep{Milgrom2010}. The latter formulation is quasi-linear. The modified Poisson's equation reads
\beq\label{QUMOND} \nabla^2 \Phi =\nabla \cdot \left[\nu({\rm Y}) \nabla \Phi_{\rm N}\right],
\eeq
where Milgromian and Newtonian gravitational potentials are denoted as $\Phi$ and $\Phi_{\rm N}$, and ${{\rm Y}\equiv |\nabla \phi_{\rm N}|/a_0}$. Here $\nu({\rm Y})$ is an interpolating function. When $|\nabla \phi_{\rm N}| >> a_0$, $\nu \rightarrow 1$ and when $|\nabla \phi_{\rm N}| << a_0$, $\nu \rightarrow 1/\sqrt{\rm Y}$. The right-hand side of Eq. \ref{QUMOND} is related only to the Newtonian gravitation induced by the observed baryonic matter. In the following sections, we will adopt the QUMOND formulation to investigate the distribution of gaseous components from the left-hand side of Eq. \ref{QUMOND}.
We shall use the ``simple'' form for the $\nu$-function:
\beq \nu(\mathrm{Y}) =1/2+\sqrt{1/4+1/\mathrm{Y}}.\eeq
The Poisson equation can be reformulated as {follows:}
\bey\label{pdm} \nabla^2 \Phi = 4\pi G\left(\rho_{\rm b} +\rho_{\rm PDM}\right),\nonumber \\
\rho_{\rm PDM}=\frac{1}{4\pi G} \nabla \cdot [(\nu-1)\nabla\Phi_{\rm N}],\eey
where $\rho_{\rm b}$ represents the baryonic density and $\rho_{\rm PDM}$ denotes the density of the phantom dark matter, which is determined by $\rho_{\rm b}$. 
Considering the assumption of spherical symmetry of our mass models, other formulations of MOND, such as AQUAL, are also applicable.

\subsection{ESD fitting method and results}
\label{ESDfit}
\subsubsection{GGL Method}
In the framework of QUMOND theory, the $\rho_{\rm b} (r)$ and $\rho_{\rm PDM}(r)$ profiles can be calculated from the modified Poisson's equation, Eq. \ref{pdm}. In spherically symmetric coordinates, the radial direction is segmented with scale intervals of $0.1\kpc$ within the radius of $R(i)\in (0,~2000]~\kpc$ and the polar direction is divided with angular intervals of 0.01 rad within the angle of $\theta(j) \in [-1.57,~1.57] $ rad. Once the baryonic density $\rho_{\rm b}$ with undetermined parameters is given, the distribution of $\rho_{\rm PDM}$ can be derived from Eq. \ref{pdm}. 

In a galaxy, $\Sigma(R)$ can be obtained by integrating the 3D density profile $\rho_{\rm b}$ along the LoS direction using Simpson's method, with the gas parameters $\Sigma_0,R_{\rm s},\rho_0,r_{\rm c}$ and $\gamma$ to be determined
\bey \Sigma(R) &=& \int_{-\infty}^{\infty} \left[\rho_{\rm b} (r) +\rho_{\rm PDM}(r)\right] dz \nonumber \\
&=& R\int_{-\pi/2}^{\pi/2} \frac{\rho_{\rm b} (R,\theta)+\rho_{\rm PDM}(R,\theta)}{\cos^2\theta} d\theta .\eey
The ESD profile of a lens central galaxy is defined as
\begin{equation}\label{esd}
    \Delta\Sigma(R)\equiv \Sigma(R)-\bar\Sigma(R)=\Sigma(R)-\frac{2}{R^2}\int_0^R\Sigma(R')R'\mathrm{d}R'{,}
\end{equation}
where $\Sigma(R)$ is the projected surface density of the galaxy at a projected radius $R$ {and} $\bar\Sigma(R)$ is the average surface density within $R$. 

With the above method, the ESD profiles can be described as a function of the model parameters, \beq \Delta\Sigma(R)=f(R,M_*,R_e,\Sigma_0,R_{\rm s},\rho_0,r_{\rm c},\gamma).\eeq
For each sub-sample of galaxies, we numerically fit their parameterized ESD profiles in MOND using a Bayesian method combined with MCMC sampling. We sample the posterior distribution of the parameters with the MCMC sampler \textit{emcee} \citep{foreman-mackeyEmceeMCMCHammer2013}. 
Moreover, {the} Bayesian estimation of the uncertainty of a derived quantity such as $M_{\rm g}$ is calculated from each set of sampled parameters.

\begin{table*}
    \renewcommand{\arraystretch}{1.3}
    \tabcolsep=0.015cm
    \centering
    \caption{The properties of star-forming and total galaxy samples.
    Twelve properties, including $^\mathrm{(a)}$the stellar mass range, 
    $^\mathrm{(b)}$the mean stellar mass, 
    $^\mathrm{(c)}$the gas mass $\log_{{10}}{M_{\rm g}/M_\odot}$ obtained from the GGL method, 
    $^\mathrm{(d)}$the {stellar fraction} $M_*/(M_* + M_{\rm g})$ obtained from the GGL method,
    $^\mathrm{(e)}$the central surface density $\Sigma_0$ of HI gas in the unit of $10^{10}M_\odot\ \mathrm{kpc^{-2}}$,
    $^\mathrm{(f)}$the scale length $R_{\rm s}$ of HI gas in the unit of $\mathrm{kpc}$,
    $^\mathrm{(g)}$the central density $\rho_0$ of the hot gas halo in the unit of $10^{5}M_\odot\ \mathrm{kpc^{-3}}$,
    $^\mathrm{(h)}$the core radius $r_{\rm c}$ of the hot gas halo profile in the unit of $\mathrm{kpc}$,
    $^\mathrm{(i)}$the power law index $\gamma $ of the hot gas halo profile obtained from the GGL method,
    $^\mathrm{(j)}$the gas mass $\log_{{10}}{M_{\rm g}/M_\odot}$ obtained from the SK method,
    $^\mathrm{(k)}$the {stellar fraction} $M_*/M_{\rm b}$ obtained from the SK method,
    $^\mathrm{(l)}$the constant anisotropy parameter $\beta$ obtained from the SK method, 
    of each sample are listed in each column, respectively.
    The errors for each property of the 12 sub-samples indicate the {16\textsuperscript{th}} and {84\textsuperscript{th}} percentiles of the posterior distribution of the property.}
    \begin{tabular}{ccccccccccccc}
    \hline
        Sub- & $\log M_*$-range$^\mathrm{(a)}$ & $\log{M_*}$$^\mathrm{(b)}$ & $\log M_{\rm g,GGL}$$^\mathrm{(c)}$ & ${f_{\star,\rm GGL}}$$^\mathrm{(d)}$ & $\Sigma_0$$^\mathrm{(e)}$ & $R_{\rm s}$$^\mathrm{(f)}$ & $\rho_0$$^\mathrm{(g)}$ & $r_{\rm c}$$^\mathrm{(h)}$ & $\gamma$$^\mathrm{(i)}$ & $\log M_{\rm g,SK}$$^\mathrm{(j)}$ & ${f_{\star,\rm SK}}$ $^\mathrm{(k)}$ & $\beta$ $^\mathrm{(l)}$\\
        sample & $(\Msun$)& $(\Msun$) & $(\Msun$) &- & $(10^{10}\msun/\kpc^2)$ &($\kpc$) & $(10^5\msun/\kpc^3)$ & $(\kpc )$ & -& $(\msun)$ &- & -\\
    \hline
    SF 1 & [8.84, 9.34] & 9.14 & $9.52_{-0.48}^{+0.41}$ & $0.29_{-0.15}^{+0.26}$ & $0.03_{-0.03}^{+0.06}$ & $0.39_{-0.27}^{+0.78}$ & $0.13_{-0.11}^{+0.43}$ & $38.60_{-12.66}^{+8.38}$ & $1.62_{-0.08}^{+0.03}$ & $10.51_{-0.45}^{+0.25}$ & $0.04_{-0.02}^{+0.07}$ & $0.26_{-0.18}^{+0.17}$\\
    SF 2 & [9.34, 9.84] & 9.64 & $10.03_{-0.35}^{+0.26}$ & $0.29_{-0.11}^{+0.19}$ & $0.10_{-0.09}^{+0.21}$ & $0.43_{-0.26}^{+1.09}$ & $0.18_{-0.13}^{+0.65}$ & $48.78_{-17.39}^{+12.17}$ & $1.62_{-0.10}^{+0.04}$ & $9.91_{-0.44}^{+0.25}$ & $0.35_{-0.12}^{+0.25}$ & $0.23_{-0.16}^{+0.19}$\\
    SF 3 & [9.84, 10.34] & 10.13 & $9.76_{-0.43}^{+0.29}$ & $0.70_{-0.16}^{+0.16}$ & $0.37_{-0.34}^{+0.63}$ & $0.35_{-0.23}^{+0.95}$ & $0.02_{-0.02}^{+0.08}$ & $61.54_{-21.55}^{+14.13}$ & $1.61_{-0.07}^{+0.04}$ & $10.44_{-0.40}^{+0.22}$ & $0.33_{-0.10}^{+0.22}$ & $0.09_{-0.06}^{+0.11}$\\
    SF 4 & [10.34, 10.84] & 10.61 & $10.45_{-0.25}^{+0.14}$ & $0.59_{-0.08}^{+0.13}$ & $0.82_{-0.74}^{+1.74}$ & $0.62_{-0.29}^{+1.26}$ & $0.03_{-0.03}^{+0.20}$ & $75.31_{-34.90}^{+18.84}$ & $1.62_{-0.08}^{+0.03}$ & $10.58_{-0.48}^{+0.24}$ & $0.52_{-0.13}^{+0.25}$ & $0.08_{-0.06}^{+0.13}$\\
    SF 5 & [10.84, 11.34] & 11.03 & $10.55_{-0.37}^{+0.26}$ & $0.75_{-0.13}^{+0.12}$ & $2.18_{-2.07}^{+5.53}$ & $0.32_{-0.22}^{+1.29}$ & $0.04_{-0.03}^{+0.08}$ & $103.79_{-32.56}^{+19.10}$ & $1.62_{-0.07}^{+0.03}$ & $10.50_{- nan}^{+0.48}$ & $0.77_{-0.24}^{+0.44}$ & $0.12_{-0.08}^{+0.15}$\\
    SF 6 & [11.34, 11.82] & 11.43 & $11.53_{-0.40}^{+0.30}$ & $0.44_{-0.16}^{+0.22}$ & $4.32_{-4.09}^{+13.64}$ & $0.49_{-0.32}^{+1.84}$ & $0.29_{-0.22}^{+0.97}$ & $128.35_{-50.26}^{+31.25}$ & $1.62_{-0.06}^{+0.03}$ & $12.06_{-1.68}^{+0.81}$ & $0.19_{-0.15}^{+0.73}$ & $0.27_{-0.18}^{+0.16}$\\
    Total 1 & [8.84, 9.34] & 9.14 & $9.84_{-0.49}^{+0.40}$ & $0.17_{-0.09}^{+0.21}$ & $0.04_{-0.03}^{+0.06}$ & $0.38_{-0.27}^{+0.97}$ & $0.29_{-0.21}^{+1.01}$ & $38.75_{-13.50}^{+8.03}$ & $1.61_{-0.09}^{+0.04}$ & $10.56_{-0.36}^{+0.19}$ & $0.04_{-0.01}^{+0.04}$ & $0.26_{-0.18}^{+0.15}$\\
    Total 2 & [9.34, 9.84] & 9.64 & $10.29_{-0.31}^{+0.22}$ & $0.18_{-0.06}^{+0.13}$ & $0.08_{-0.07}^{+0.19}$ & $0.50_{-0.33}^{+1.02}$ & $0.43_{-0.28}^{+1.89}$ & $46.58_{-19.59}^{+11.61}$ & $1.62_{-0.07}^{+0.03}$ & $10.29_{-0.21}^{+0.19}$ & $0.18_{-0.06}^{+0.08}$ & $0.16_{-0.12}^{+0.17}$\\
    Total 3 & [9.84, 10.34] & 10.15 & $10.27_{-0.20}^{+0.19}$ & $0.43_{-0.10}^{+0.11}$ & $0.44_{-0.41}^{+0.64}$ & $0.52_{-0.23}^{+1.44}$ & $0.19_{-0.15}^{+4.23}$ & $47.65_{-28.05}^{+18.25}$ & $1.62_{-0.07}^{+0.04}$ & $10.62_{-0.32}^{+0.19}$ & $0.25_{-0.08}^{+0.16}$ & $0.09_{-0.07}^{+0.13}$\\
    Total 4 & [10.34, 10.84] & 10.64 & $10.92_{-0.07}^{+0.06}$ & $0.35_{-0.03}^{+0.03}$ & $0.41_{-0.36}^{+1.66}$ & $0.96_{-0.64}^{+1.84}$ & $54.01_{-51.09}^{+459.38}$ & $13.03_{-6.88}^{+18.73}$ & $1.60_{-0.10}^{+0.05}$ & $11.02_{-0.20}^{+0.15}$ & $0.29_{-0.07}^{+0.10}$ & $0.05_{-0.04}^{+0.08}$\\
    Total 5 & [10.84, 11.34] & 11.09 & $11.45_{-0.04}^{+0.04}$ & $0.30_{-0.02}^{+0.02}$ & $0.64_{-0.59}^{+3.15}$ & $1.24_{-0.77}^{+3.51}$ & $37.38_{-30.44}^{+94.65}$ & $22.54_{-7.77}^{+15.39}$ & $1.62_{-0.09}^{+0.04}$ & $11.58_{-0.14}^{+0.11}$ & $0.25_{-0.04}^{+0.06}$ & $0.03_{-0.02}^{+0.04}$\\
    Total 6 & [11.34, 11.84] & 11.47 & $12.63_{-0.03}^{+0.03}$ & ${0.06_{-0.004}^{+0.005}}$ & $0.10_{-0.09}^{+0.58}$ & $4.53_{-2.94}^{+13.88}$ & $15.92_{-4.59}^{+6.24}$ & $77.68_{-14.73}^{+13.40}$ & $1.54_{-0.18}^{+0.07}$ & $12.53_{-0.10}^{+0.08}$ & $0.08_{-0.01}^{+0.02}$ & $0.06_{-0.04}^{+0.08}$\\
    \hline
    \end{tabular}
    \label{tab:samples}
\end{table*}

\begin{figure*}
    \centering
    \includegraphics[width=0.75\textwidth]{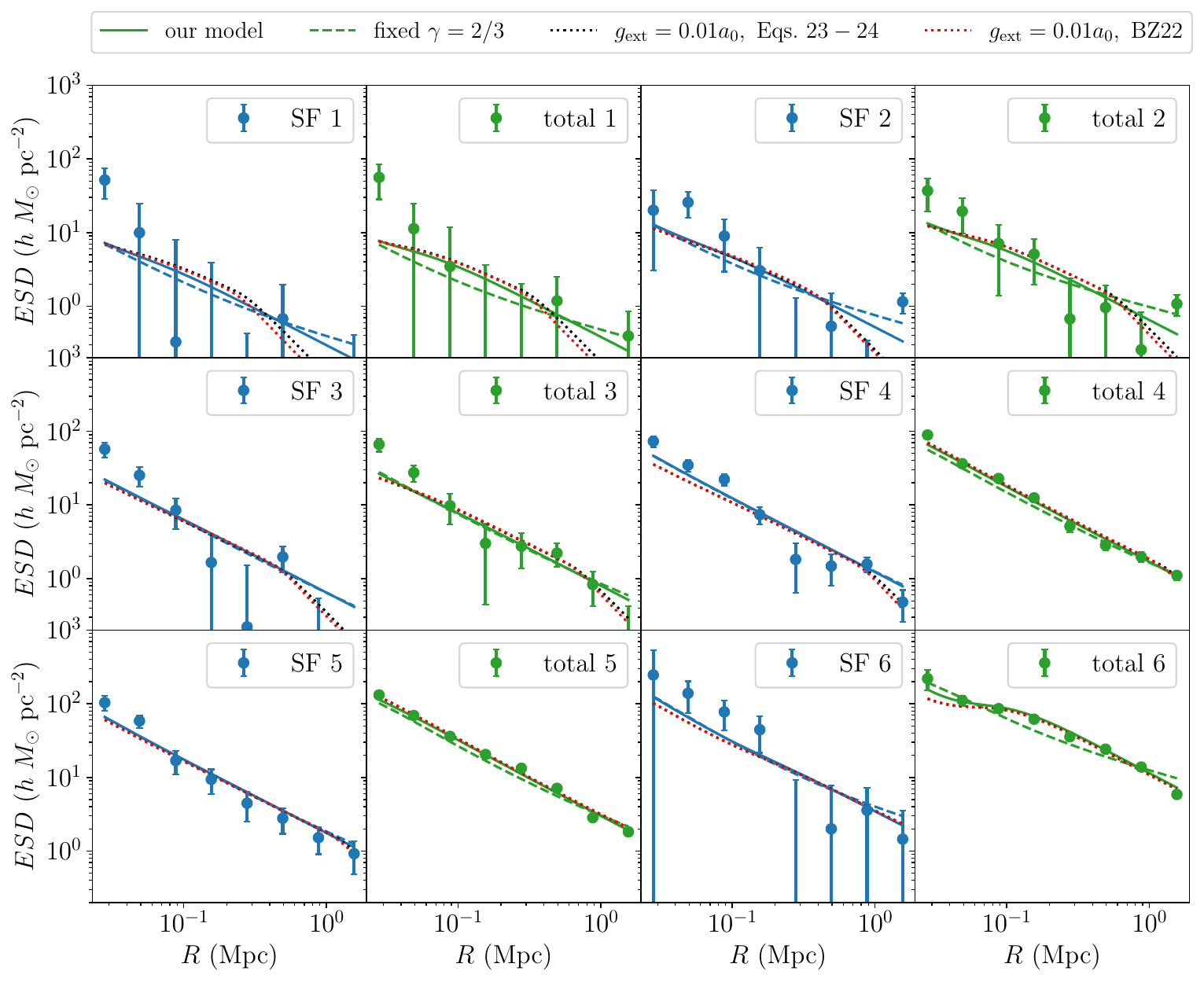}
    \caption{The ESD profiles of the galaxy samples and the best fitting results using the three-component baryonic mass model.
    The blue and green symbols show the ESD profiles of the star-forming galaxies (first and third columns) and the total galaxy samples (second and fourth columns).
    The ESD profiles obtained from the best-fitting models, 
    including our model (solid), the isothermal hot gas halo model (dashed), and the model with the existence of an external field ({black dotted for our analytic formulations and red dotted for BZ22}),
    are shown by the blue and green lines for different galaxy samples.}
    \label{fig:ESD_disp}
\end{figure*}

\subsubsection{Profile parameters}
The best-fitting profiles are shown in solid curves in Fig. \ref{fig:ESD_disp}. The observed ESD profiles for 12 sub-samples of galaxies are compared with the theoretically predicted ESD profiles in Fig. \ref{fig:ESD_disp}. We find that the ESD profiles predicted in MOND agree well with the observations, especially in the outer regions of the galaxies.
Moreover, the parameters of the profiles of the cold and hot gas halos are determined. We present the gas profile parameters in the $6^{{th}}$-$10^{{th}}$ columns in Table \ref{tab:samples}. 

For the cold gas halos, in the sample of star-forming galaxies, the central surface density ($\Sigma_0$) of the cold gas component exhibits an increase with the growth of stellar mass. The scale of the cold gas halo ($R_{\rm s}$) barely varies with changes in the stellar mass. In the total sample, the fitted parameters for cold gas density profiles exhibit similar trends to the stellar masses, except in the sub-sample Total 6.
The exception indicates that the cold gas of this sub-sample is rather diffuse.  

Our main focus is the hot gas halos. In the sample of star-forming galaxies, the core radius ($r_{\rm c}$) increases as the stellar mass grows. 
Unlike the star-forming sample, $r_{\rm c}$ of the hot gas in the total sample does not monotonically increase with the growth of $M_*$. However, the value of $\gamma$ is almost a constant, $\approx 1.62$, for both samples of galaxies. This leads to $\rho_{\rm hg} \propto r^ {-4.8}$, which is close to a Plummer model \citep{plummer_problem_1911} where $\rho\propto (1+(r^2/r_p^2))^{-5/2} \propto r^{-5}$. The exception is in the sub-sample of Total 6, where $\gamma \simeq 1.54$. The marginally lower value of $\gamma$ in this sub-sample implies a somewhat more extended hot gas distribution in the outer regions. However, the difference in the density power law index is not {significant}. 

In the X-ray observations, the hot gas halo around the Milky Way is diffuse and extended, with $\gamma=0.64-0.68$ and peaked at $\gamma=2/3$ \citep{nicastroXRayDetectionGalaxy2023}, which indicates a much more extended hot gas halo within $115 ~\kpc$, and $\rho_{\rm X-ray} \propto r^{-2}$ when $r>> r_c$. {As mentioned above, our MOND predicted gas halo has a more sharply declining profile outside the core radius, $\rho_{\rm hg} \propto r^{-4.8}$, closely approximating the Plummer distribution at large distances $\propto r^{-5}$.} However, it is still too early to conclude that MOND has been ruled out. For comparison, we perform the ESD fitting again using an isothermal hot gas halo model (i.e., a fixed $\gamma=2/3$) and show the results in dashed curves in Fig. \ref{fig:ESD_disp}. We find that the isothermal hot gas halos also predict ESD profiles that agree well with most sub-samples of observations within the $1\sigma$ error range, including the sub-samples with similar stellar mass {to} the Milky Way. In some sub-samples, we cannot distinguish the fitted ESD profiles from a Plummer-like to an isothermal gas halo, such as sub-sample SF 3-6, and Total 3. A possible reason for the MOND predicted values of $\gamma$ significantly differing from the observations is that the large uncertainties in the observed ESD data allow for a wide range of $\gamma$ values. In several sub-samples with small observational errors, particularly in total 6, our sharply declining $\gamma=1.6$ models exhibit better agreement with the data than the $\gamma=2/3$ models. In addition, another possible reason could be that the virial radius in MOND is much further {beyond the observed $115~\kpc$ of the X-ray study by \citet{nicastroXRayDetectionGalaxy2023}.} We fit the ESD profiles within $115\kpc$ of the 12 sub-samples, and the density power law indices become significantly smaller. For instance, in sub-samples such as SF 2, Total 2, and Total 3, $\rho_{\rm hg} \propto r^{-2}$ within this truncation radius. This suggests that the distribution of hot gas halos is relatively extended within $100 ~\kpc$, while the density rapidly decreases at larger radii. Our prediction offers a new probe to test gravity. Future observations on the distribution of X-ray gas halos around central galaxies will provide a valuable comparison.

\subsubsection{Gas masses and {stellar fraction}}
Once the parameters for the $\rho_{\rm b}(r)$ profiles are determined, the probability distribution of the gas mass $M_{\rm g}$ can be derived from the posterior distribution of the model parameters. The mass of the gaseous component is defined as the total mass of HI gas and the hot gas halo within the virial radius, $r_\mathrm{vir}$, of the central galaxy.
The relation between the virial mass of the phantom dark matter halo plus the baryonic mass of a galaxy in the MOND framework can be expressed as \citep{wuSpecificFrequencyGlobular2013}
\beq
    M_\mathrm{vir} \equiv  pr_\mathrm{vir}^3 =v_{\rm c}^2r_{\rm vir}/G,
\eeq
with $v_{\rm c}$ being the circular velocity at the infinity radius of a galaxy. In MOND dominated regimes, $v_{\rm c} = (GM_{\rm b} a_0)^{1/4}${,} where $M_{\rm b}=M_* + M_{\rm g}$ is the total mass of baryons in a galaxy.
Thus the virial mass of a central galaxy reads 
\beq
     M_\mathrm{vir} = (Ga_0M_{\rm b})^{3/4}p^{-1/2}G^{-3/2},
\eeq
where $p=\frac{4}{3}\pi \times 200\rho_\mathrm{crit}$. $\rho_\mathrm{crit}=\frac{3H^2}{8\pi G}$ is the critical density of the Universe with a Hubble constant of $H=67.64\ \mathrm{km\ s^{-1}\ Mpc^{-1}}$ {today \citep{Tristram+2024}.} Thus $M_{\rm g}$ and $r_\mathrm{vir}$ are calculated iteratively by solving Eq. \ref{QUMOND}. 

\begin{figure*}
    \centering
    \includegraphics[width=1.0\textwidth]{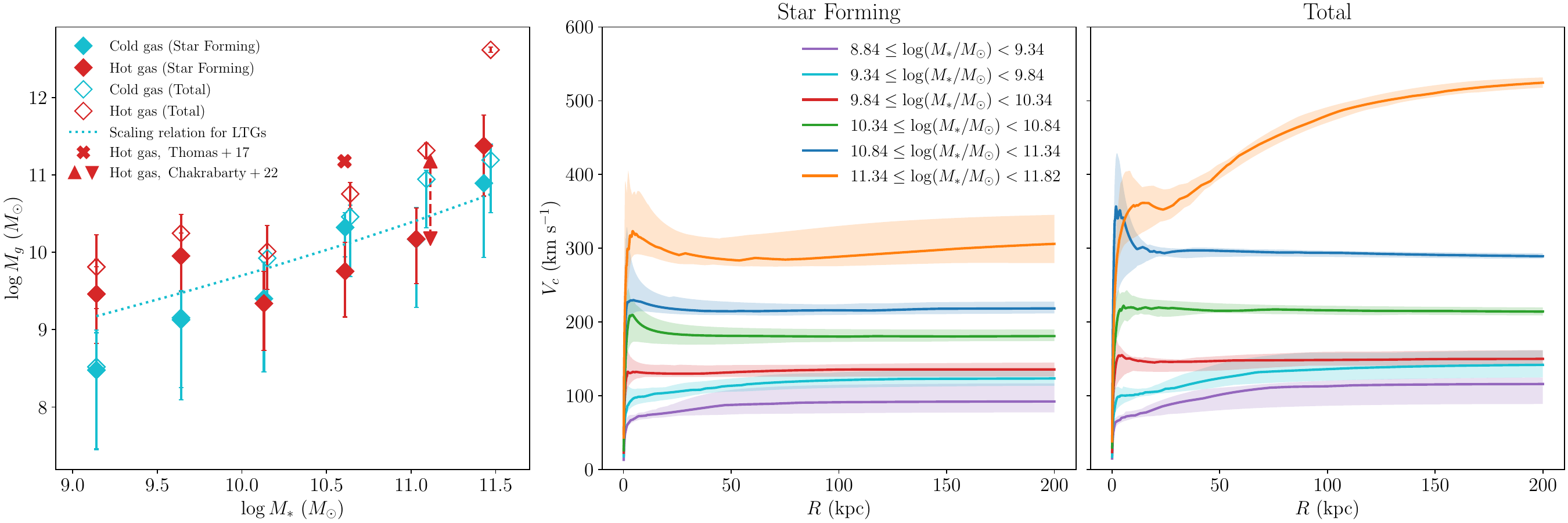}
    \caption{Left panel: The gas mass estimation from GGL method. The symbols represent the estimated cold gas mass in star-forming galaxies (cyan, filled) and in total galaxies (cyan, open), and the estimated hot gas mass in star-forming galaxies (red, filled) and total galaxies (red, open). The error bars of the symbols represent the {16\textsuperscript{th}} and {84\textsuperscript{th}} percentiles of the posterior distribution. The scaling relations of cold gas mass in star-forming {galaxies} (cyan, dotted) in \citet{Mistele+2024} is shown for comparison. The hot gas mass of the Milky Way in the models of \citet{thomas2017a} (red X symbol) and \citet{chakrabarty2022} (red upward triangle for the maximal value and red downward triangle for the minimal value) are also displayed. Middle to right panels: Rotation curves calculated within { the MOND framework} for the sub-samples of star-forming and total galaxies. The colored {shaded} areas represent the $1\sigma$ confidence interval.}
    \label{rc}
\end{figure*}

We listed the median values of gas masses and the corresponding errors of the 12 sub-samples of galaxies in the $4^{{th}}$ columns of Table \ref{tab:samples}. 
The calculated masses of cold and hot gas in all sub-samples are shown in the left panel of Fig. \ref{rc}. In general, $M_{\rm g}$ increases as the stellar masses grow in both samples of galaxies. Our mass estimation of cold gas in star-forming galaxies is consistent with the scaling relation of cold gas in star-forming late-type galaxies in \citet{Mistele+2024}. 

We also compare the estimated hot gas mass in the existing Milky Way QUMOND models proposed by \citet{thomas2017a} and \citet{chakrabarty2022}, in the left panel of Fig. \ref{rc}. \citet{thomas2017a} used an oblate hot gas halo model with a minor-to-major axial ratio of $a_{\rm HG}=0.44$. Their predicted hot gas mass is significantly higher than our results at a similar stellar mass. However, in a more sophisticated model suggested by \citet{chakrabarty2022}, two values for $a_{{\rm hg}}$ are considered, 0.4 and 0.8, respectively. Our estimated hot gas mass falls within their predicted range.

In addition, {the stellar fraction}, denoted as $\conveff$, can be calculated by $\conveff = M_*/(M_* + M_{\rm g})$. The values of $\conveff$ for the 12 sub-samples computed from the GGL are presented in the $5^{{th}}$ column of Table \ref{tab:samples}. 
The values of $\conveff$ here differ from those {gas-to-star conversion} efficiency in \cite{Zhang+2022}, where the baryonic mass of a central galaxy is defined as a product of a constant cosmic fraction $f_{\rm b}$ and the dark matter halo mass $M_{\rm h}$. 
{As previously mentioned, our baryonic mass estimates are derived using weak gravitational lensing in conjunction with MOND dynamics, providing more accurate baryonic mass measurements for central galaxies within each mass bin.  This approach eliminates errors in baryonic mass estimation caused by individual galaxies deviating from the constant value of $f_{\rm b}$.} 
We find that the $\conveff$ values for the star-forming sample of galaxies are higher than those for the total sample at the same stellar mass{, indicating a higher gas-to-star conversion efficiency}. Such a conclusion {qualitatively} agrees well with that in \cite{Zhang+2022}.

\subsubsection{Rotation curves}
The rotation curves of the sub-samples of star-forming and total galaxies can be calculated from the GGL-fitted baryonic density, $\rho_{\rm b}$. We demonstrate the rotation curves in the middle and right panels of Fig. \ref{rc}. The $1\sigma$ confidence interval is induced by the uncertainties of the five parameters in the ESD fitting, computed using a Monte Carlo method. 

The rotation curves are almost flat from a few tens of kilo-parsec for all the star-forming sub-samples and most of the total sub-samples. The flat rotation curves in large radii are in good agreement with the observations \citep[e.g.,][]{Mistele+2024}. There is one exception, namely sub-sample Total 6. The rotation curve in this sub-sample is mildly rising out to 100 $\kpc$. However, within the radial range of 100-200 $\kpc$, the rotation curve becomes asymptotically flat. The reason for the rising rotation curve within 100 $\kpc$ in sub-sample Total 6 might be that the gas distribution in this sub-sample is more extended compared to other sub-samples, as {mentioned above}.
Moreover, $\conveff {=0.06_{-0.004}^{+0.005}} $ for this sub-sample, indicating a gas-dominated system.

\section{SK for determining galactic baryonic mass}
\label{sk}
Aside from gravitational lensing, SK can also probe into the total mass distribution of central galaxies.
In the deep MOND regime, a simple formula \beq\label{isotropic}\sigma^2_\mathrm{LoS}=\frac{2}{9}(GM_{\rm b}a_0)^{1/2}\eeq exists for the LoS dispersion of an isolated low surface density system.
A modified formula is proposed \citep{haghiNewFormulationExternal2019b} in the presence of an external field.
In a more realistic context, the LoS velocity dispersion profiles of satellites, $\sigma_{\rm LoS}$, are anisotropic. The anisotropic velocity profiles have been introduced in both MOND and CDM frameworks to compare with the observations
\citep{lokasVelocityDispersionsDwarf2001,angusVelocityDistributionSloan2007,tiretVelocityDispersionEllipticals2007,angusDwarfSpheroidalsMOND2008,klypinTESTINGGRAVITYMOTION2009}. The observations revealed that anisotropic velocity dispersion is required to agree with the observed kinematics of the satellites in both frameworks. Here we will begin with constant anisotropies for the satellites around their central galaxies in the 12 sub-samples.

\begin{figure*}
    \centering
    \includegraphics[width=\textwidth]{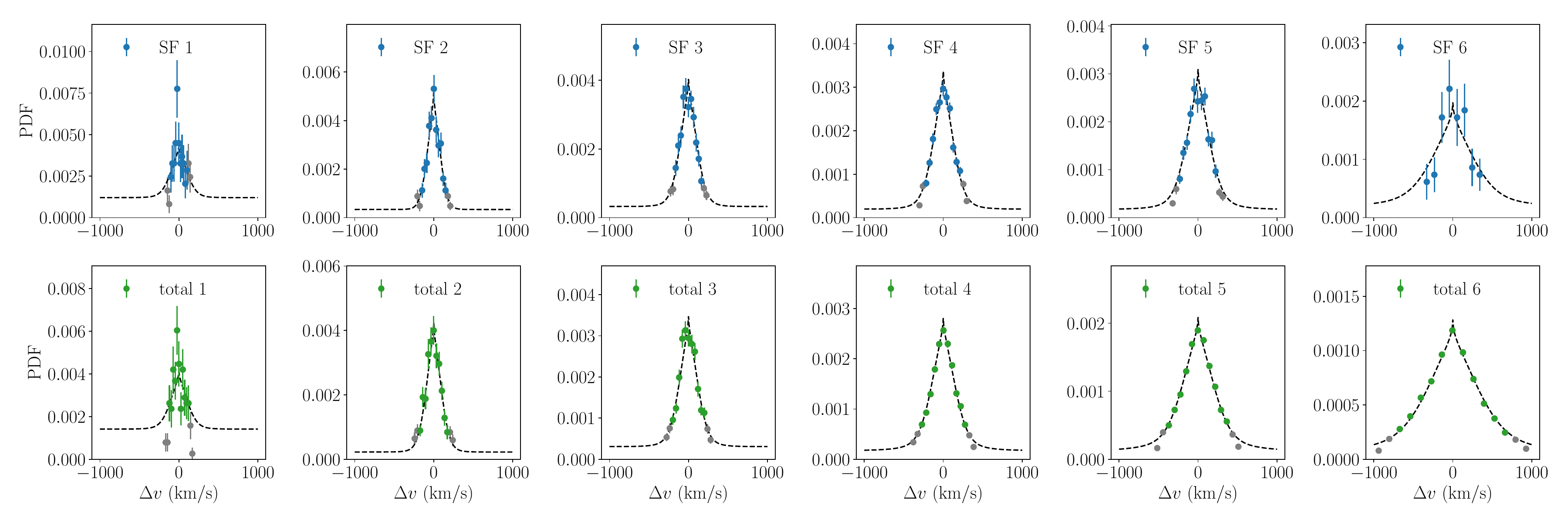}
    \caption{The PDFs of the LoS velocity difference of satellite galaxies and the best fitting results.
    The blue and green symbols with error bars represent the PDFs of satellites around the star-forming galaxies (upper row), and around the total galaxies (lower row), respectively.
    The grey symbols represent the data points excluded by the $|\Delta v|<1.7v_\mathrm{200}$ cut.
    The dotted lines represent the best-fitting results of the PDFs using our modified Gaussian model in Eq. \ref{mod_Gaussian}.
    The results of star-forming and total samples in different stellar mass bins are shown in different columns.}
    \label{fig:SKfit}
\end{figure*}

\subsection{Method to calculate the LoS velocity dispersion}\label{skmethod}
There are some pioneering works on testing MOND by the $\sigma_{\rm LoS}$ profiles of satellite galaxies \citep{angusVelocityDistributionSloan2007,klypinTESTINGGRAVITYMOTION2009}. We will follow their method to calculate the $\sigma_{\rm LoS}$ profiles of the satellites in the framework of MOND. We need to solve the Jeans equation. Under the assumption of spherical symmetry, the Jeans equation can be written in the form of
\begin{equation}\label{jeans}
    \frac{\mathrm{d}\sigma^2}{\mathrm{d}r}+\sigma^2\frac{2\beta+\alpha}{r}=-g(r),
\end{equation}
where $\sigma(r)$ is the radial velocity dispersion, $\beta=1-\sigma_\bot^2/2\sigma^2$ is the anisotropy parameter, 
$\alpha=|\mathrm{d}\ln \nden (r)/\mathrm{d}\ln r|$ is the slope of the satellite number density, $\nden (r)$. Assuming the satellites are test particles, the general solution of the Jeans equation can be expressed as \citep{angusVelocityDistributionSloan2007,klypinTESTINGGRAVITYMOTION2009}
\begin{equation}\label{sigma2}
    \sigma^2(r)=\frac{1}{\chi(r)\nden (r)}\int_r^\infty \chi(r)\nden (r)g(r)\mathrm{d}r,
\end{equation}
where $\chi(r)=\exp\left[2\int_0^r \beta(r)\right]r^{-1}\mathrm{d}r$. The Milgromian gravitational acceleration $g(r)$ here is calculated by ${\bf g(r)}=-{\bf \nabla}\Phi$ from Eq. \ref{QUMOND}, and approximately the Newtonian acceleration $g_{\rm N} =GM_{\rm b}/r^2$ with the central galaxy being a point mass particle. Such an approximation is reasonable since the distances of satellites are much larger than the typical size of central galaxies. 
The $\sigma_{\rm LoS}$ profile can be obtained by integrating $\sigma(r)$ along the LoS direction. The projected pressure of a cluster of galaxies can be written as \citep[Eq. A7 in][]{Mamon_Lokas2005}
\beq \label{pressure}
\Sigma_\nden(R) \sigma_{\rm LoS}^2(R) = 2\int_{R}^{\infty}\left(1-\beta \frac{R^2}{r^2}\right)\nden (r)\sigma^2(r) \frac{r{\rm d}r}{\sqrt{r^2-R^2}},
\eeq
where $\Sigma_\nden(R)$ is the projected number density of the satellites along the LoS. Now we can derive the $\sigma_{\rm LoS}$ profile from Eqs. \ref{sigma2}-\ref{pressure} with the undetermined $M_{\rm b}$.

\subsection{Baryonic mass determined by SK}\label{skvalues}
\begin{figure*}
    \centering
    \includegraphics[width=\textwidth]{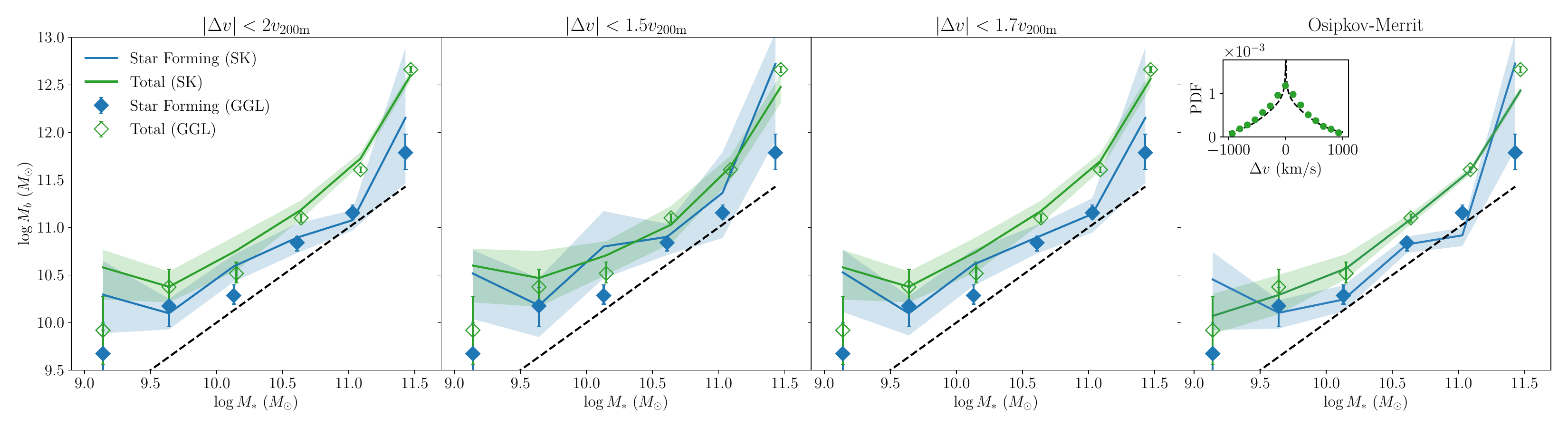}
    \caption{The baryonic mass estimation from GGL and SK.
    In each panel, the symbols represent the baryonic mass estimation of the SF (blue, filled) and total (green, open) galaxy samples from the GGL method.
    The error bars of the symbols represent the {16\textsuperscript{th}} and {84\textsuperscript{th}} percentiles of the posterior distribution.
    The solid lines represent the baryonic mass estimation of the SF (blue) and total (green) galaxy samples from the SK method.
    The shaded regions represent the {16\textsuperscript{th}} and {84\textsuperscript{th}} percentiles of the posterior distribution.
    The first three panels represent the results for different cut values: $|\Delta v|<2v_\mathrm{200}$, $|\Delta v|<1.5v_\mathrm{200}$, $|\Delta v|<1.7v_\mathrm{200}$, respectively. The rightmost panel displays the mass relation when the $\sigma_{\rm LoS}$ of satellites follows the Osipkov-Merritt anisotropy.}
    \label{fig:Ms-Mb}
\end{figure*}

\subsubsection{Method to fit the PDFs of relative velocities}\label{fitPDFs}
The baryonic mass of each galaxy sub-sample can be estimated from the PDFs of the LoS velocity difference of the satellites.  
We used the PDFs of $\Delta v$ of the satellites for the 12 sub-samples in \cite{Zhang+2022}. The fitting for the PDFs of $\Delta v$ follows
\begin{equation}
    \frac{A}{\sqrt{2\pi}\sigma_\mathrm{LoS}}\Sigma_\nden(R) \exp\left[-\frac{\Delta v^2}{2\sigma^2_\mathrm{LoS}(R)}\right]+d.
    \label{mod_Gaussian}
\end{equation}
Here the Gaussian term with variable dispersion stands for the real satellites, and the constant term $d$ represents the foreground interlopers which are not gravitationally bound to the cluster. In the previous work in the framework of $\Lambda$CDM \citep{Zhang+2022}, they selected satellites around the central galaxies following the criteria of $|\Delta v|\leq 3v_\mathrm{200}$, $R_\mathrm{p}\leq r_\mathrm{200}$ and $M_s<M_*$.
Here, $\Delta v$ and $R_\mathrm{p}$ are the LoS relative velocity and the relative projected distance between the satellite and the central galaxy, respectively. The circular velocity at the virial radius, $r_{\rm 200}$, of the Navarro-Frenk-White (NFW; \citealp{navarroUniversalDensityProfile1997a}) dark matter halo, derived from lensing data, is denoted as $v_\mathrm{200}$. 

However, note that the criterion $|\Delta v|\leq 3v_\mathrm{200}$ in \cite{Zhang+2022} is not strict enough to exclude all the interlopers, as further pointed out by \cite{Zhang+2024b}.
A new criterion $|\Delta v|< |\Delta v|_\mathrm{cut}$ was suggested to reduce the contamination of the interlopers.
Stable fitting results were achieved with $|\Delta v|_\mathrm{cut}$ in the range of $1.3v_\mathrm{200}$ to $1.8v_\mathrm{200}$ \citep{Zhang+2024b}. 

In this work, we test three different values of $|\Delta v|_{\rm cut}$: $|\Delta v|_{\rm cut} = 2v_\mathrm{200}, ~1.5v_\mathrm{200}$ and $1.7v_\mathrm{200}$. We take the truncation of $|\Delta v|_{\rm cut}$ at $1.7v_\mathrm{200}$ for the PDFs of the LoS velocity difference of the satellites as an example. The PDFs of $\Delta v$ can be expressed as a function of 3 parameters $p[\Delta v|\vec{\theta}(M_{\rm b},\alpha,\beta)]$. We use the MCMC sampler \textit{emcee} again to sample the posterior distribution of the baryonic mass, $M_{\rm b}$, and the anisotropy parameter, $\beta$, with a fixed satellite number density slope $\alpha=2$ obtained from observations \citep{Jiang+2012}. The best fitting results to the PDFs of $\Delta v$ are shown in Fig. \ref{fig:SKfit}. The fitted PDFs go through precisely the observed data points except for three sub-samples, SF 1, Total 1, and Total 6, due to the relatively poor quality of data in these three sub-samples.
With the best fit to the PDFs of $\Delta v$, we obtain the mass of the gaseous component using the method of SK, denoted as $M_{\rm g, SK} = M_{\rm b} -M_*$ and shown in the $11^{{th}}$ column of Table \ref{tab:samples}. The {stellar fraction} can also be calculated using this method and is listed in the $12^{{th}}$ column of Table \ref{tab:samples}. 

{The stellar fraction} $\conveff$ calculated using SK is generally higher for the star-forming galaxies than the total galaxies in all mass bins. Similar to the results obtained from the GGL method, the global trend is that $\conveff$ peaks around $M_*=10^{10.5}M_\odot$ and decreases towards both the low and high mass ends, which agrees with \cite{Zhang+2022}.

Furthermore, the values of $\beta$ are displayed in the $13^{{th}}$ column of Table \ref{tab:samples}. 
The 12 sub-samples show mildly radial anisotropy. The galaxies are nearly isotropic or mildly radially anisotropic for the total sample of galaxies with $M_* > 10^{10.34}\msun$, where $\beta \leq 0.06$. The star-forming galaxies are slightly more radially anisotropic within the same stellar mass range, $\beta \in [0.08,~0.27]$. Our results are different from the earlier theoretical works by \citet{angusVelocityDistributionSloan2007,klypinTESTINGGRAVITYMOTION2009}, where unnaturally strongly radial anisotropy ($\beta \geq 0.6$) at $r>200~\kpc$ in MOND is required to agree with the SDSS observations. In contrast, our mildly radial anisotropic motions of satellites agree with observations in \citet{wojtakPhysicalPropertiesUnderlying2013}.

\subsubsection{Comparing gas mass obtained from two methods}\label{gasmass}
As aforementioned, the validity of the SK method largely depends on the selection of true satellites. 
In the first three panels of Fig. \ref{fig:Ms-Mb}, we show the comparison of overall baryonic mass $M_{\rm b}$ calculated from the GGL and SK methods. The baryonic mass estimated from the latter method is computed with relative velocities between the satellites and central galaxies cut in three ranges: $|\Delta v|< 2v_\mathrm{200}$, $|\Delta v|< 1.5v_\mathrm{200}$ and $|\Delta v|< 1.7v_\mathrm{200}$, following the procedure introduced in \S \ref{fitPDFs}. 
The values of the reduced chi-squared statistics between the two methods, $\chi_n^2$, within the three velocity cuts, are $6.95,~34.16$ and $9.21$, respectively. The fitting to the PDFs of relative velocities agrees better when the observations are truncated at $|\Delta v|< 1.7v_\mathrm{200}$ than at $|\Delta v|< 1.5v_\mathrm{200}$, and the results remain stable when adopting the $|\Delta v|< 2v_\mathrm{200}$ truncation although more observed data points of the PDFs of $|\Delta v|$ are included. 
Thus in the following section, the values of $M_{\rm g,SK}$ will use the ones calculated within the velocity truncation of $|\Delta v|< 1.7v_\mathrm{200}$. 

The masses of gas and $\conveff$ obtained from two independent methods, i.e., the GGL and the SK, can be cross-checked. The values from the two methods show good agreement for sub-samples with small errors in their observed ESDs. However, the deviation between $M_{\rm g, GGL}$ and $M_{\rm g, SK}$ becomes significant when the errors of the observed ESDs are large, particularly in sub-samples SF 1, SF 3, SF 6, Total 1 and Total 3. In addition, the errors of the observed PDFs are large in sub-samples of SF 1, SF 6, and Total 1. These combined errors result in discrepancies in the baryonic masses of these sub-samples when using the two methods, with deviations in the logarithmic gas mass exceeding $0.3$dex.
In the rest of the sub-samples, the deviations are less than $0.13$dex. This agreement suggests that when the quality of the observed ESDs and PDFs is sufficient, the gas mass estimation using the GGL method is reliable. However, when the errors in the observed ESDs are significant, alternative methods such as SK are necessary to verify the reasonableness of the gas mass estimation.

Furthermore, the $M_*$-$M_{\rm b}$ relation in Fig. \ref{fig:Ms-Mb} is approximately monolithic in both methods. While the gas mass predicted through GGL does not exhibit a monolithic increase with stellar mass, the overall baryonic mass does. In Fig. \ref{fig:Ms-Mb}, we find that the baryonic mass in the total sample tends to be higher than that in the star-forming sample at a given $M_*$ by using both methods, implying a lower fraction of gas in the star-forming galaxies. This is consistent with the conclusion in the $\Lambda$CDM framework \citep{Zhang+2022,Zhang+2024b}.

\section{The baryonic mass-to-LoS velocity dispersion relation}\label{msigma}
A halo-mass-to-satellite-velocity-dispersion ($M_{\rm h}-\sigma_{\rm s}$) relation was shown in \cite{Zhang+2022}. In MOND, there is no dark matter halo in a galaxy; however, a similar scaling relation is expected to exist. This scaling relation is between the baryonic mass of the central galaxy and the $\sigma_{\rm LoS}$ of the satellites, denoted as the $M_{\rm b}-\sigma_{\rm s}$ relation. In this section, we will compare our model predicted $\sigma_s$ values with the observed values obtained from \cite{Zhang+2022}.
The baryonic masses of galaxies $M_{\rm b}$ are obtained by using the method of GGL in \S \ref{GGL}. The reason for using these masses is that the three-component baryonic model for galaxies is more precise in the GGL method, especially in small radii. In contrast, a central galaxy is simplified to a point mass particle when using the SK method.

\begin{figure*}
    \centering
    \includegraphics[width=1.0\textwidth]{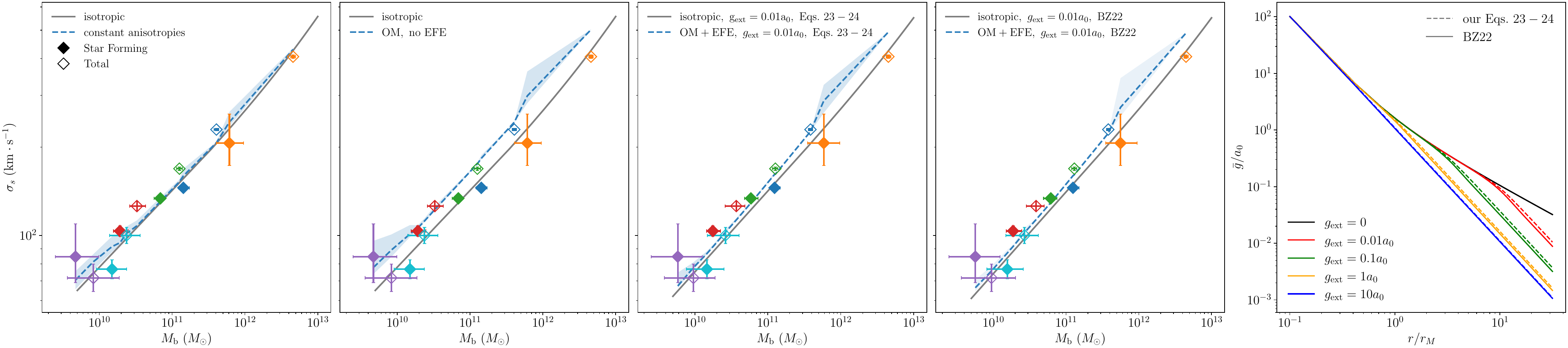}
    \caption{{The first to fourth panel:} Baryonic mass-satellite velocity dispersion relation. The open and filled symbols in the vertical axes represent observed satellite velocity dispersion for the star-forming and total galaxy samples, respectively, from \cite{Zhang+2022}. The baryonic masses in the horizontal axes are calculated from the GGL method. Different colors indicate samples of different stellar mass bins the same as those adopted in Fig. \ref{rc}. The error for $M_{\rm b}$ represents the {16\textsuperscript{th}} and {84\textsuperscript{th}} percentiles of the posterior distribution from MCMC sampling. The blue curve with {shaded region} represents the $\sigma_{\rm s}$ versus $M_{\rm b}$ predicted in MOND constant anisotropic models. 
    The Osipkov-Merritt anisotropy profiles are introduced when computing $\sigma_{\rm s}$ in the middle and right panels, incorporating an isolated environment and an external field of $0.01a_0$, respectively. Isotropic models in isolation (the {first} and {second} panels) and an external field ({the third panel for our analytic formulas and the fourth panel for BZ22}) are displayed as black curves.
    {The fifth panel: the averaged radial gravitational acceleration profiles of a point mass under different external field strengths (dashed lines for our analytic formulations and solid lines for BZ22). Accelerations are scaled in units of $a_0$ while radial distances in units of the MOND radius $r_M\equiv \sqrt{GM/a_0}$.}
   } \label{fig:M-sigma}
\end{figure*}

According to Eq. \ref{pressure}, once the form of anisotropy is given, $\sigma_{\rm LoS}(R)$ can be obtained. The mean value of $\sigma_{\rm LoS}$ for satellites, $\sigma_{\rm s}$, around a central galaxy within $r_{\rm vir}$ follows
\beq\label{lossigma}
\sigma_{\rm s}^2=\frac{\int_\delta ^{r_{\rm vir}} \Sigma_\nden (R') \sigma_{\rm Los}^2(R')\rm{d}R'}{\int_\delta ^{r_{\rm vir}} \Sigma_\nden(R') \rm{d}R'},
\eeq
where $\delta =10\kpc$.

We aim to demonstrate the relationship of $M_{\rm b}$-$\sigma_{\rm s}$ in Fig. \ref{fig:M-sigma}. In the {first} panel, initially, we present the observed, model-independent values of $\sigma_{\rm s}$ and their errors for 12 sub-samples, as reported by \citet{Zhang+2022}, fitted using the Gaussian distribution of the PDFs of $|\Delta v|$. The GGL-predicted $M_{\rm b}$ and these observed $\sigma_{\rm s}$ values suggest $M_{\rm b}\propto \sigma_{\rm s}^{3.88}$, approximately $M_{\rm b}\propto \sigma_{\rm s}^{4}$. Next, we use the SK method, using Eqs \ref{sigma2}-\ref{pressure}, to calculate the $\sigma_{\rm s}$ and the corresponding errors to the given $M_{\rm b}$\footnote{We used the same method described in \S \ref{fitPDFs}, but $M_{\rm b}$ is now fixed to the values along the horizontal axes.}. These kinematics-fitted $\sigma_{\rm s}$ and errors to the given $M_{\rm b}$ are presented with a blue curve and shadow area, respectively. The model-predicted $M_{\rm b}$-$\sigma_{\rm s}$ relationship aligns well with the model-independent observed $\sigma_{\rm s}$ versus the GGL-predicted $M_{\rm b}$ relation.

In addition, an isotropic model has been introduced into the SK method (Eqs. \ref{sigma2}-\ref{pressure}) to calculate $\sigma_{\rm s}$. This $M_{\rm b}$-$\sigma_{\rm s}$ relationship is shown in a black solid curve, which also agrees with the observed $\sigma_{\rm s}$ at the given $M_{\rm b}$.

Notably, in determining the $M_{\rm b}$-$\sigma_{\rm s}$ relationship, we assume constant anisotropies for satellites in each sub-sample and consider galaxy clusters in isolation. Despite these simplifications, the fitted relation aligns well with observations, making the simplifications reasonable.
We will further introduce more sophisticated models, including an Osipkov-Merritt anisotropy profile  \citep{osipkovSphericalSystemsGravitating1979,merrittRelaxationTidalStripping1985} and an external field for $\sigma_{\rm LoS}$ of satellites, in the following section.

\section{More sophisticated models}
\label{sophisticated}

\subsection{The Osipkov-Merritt anisotropy model}\label{OManiso}

A widely used model for the anisotropy profile to describe the motion of satellite galaxies is the Osipkov-Merritt anisotropy model \citep{osipkovSphericalSystemsGravitating1979,merrittRelaxationTidalStripping1985},
\begin{equation}
    \beta(r)=\frac{r^2}{r^2+r_a^2},
    \label{OM}
\end{equation}
where $r_a$ is the anisotropy radius. In this model, $\beta\approx 0$ when $r\ll r_a$ and $\beta\approx 1$ when $r\gg r_a$, 
representing an isotropic central region and a purely radially anisotropic outer region of a system.

Using the method described in \S \ref{skmethod}-\ref{skvalues}, with $\beta$ following the Osipkov-Merritt anisotropy profile, we can derive the radial velocity dispersion profiles of the satellites. The PDFs of $\Delta v$ can then be expressed as a function of 3 parameters $p[\Delta v|\vec{\theta}(M_{\rm b},\alpha,r_a)]$. After MCMC sampling, the baryonic mass estimates are shown in the fourth panel of Fig. \ref{fig:Ms-Mb}, with a PDF fitting result in the upper left sub-panel. Introducing Osipkov-Merritt anisotropy results in a sharp PDF peak near zero velocity due to radial satellite orbits at large radii. Generally, the masses calculated from GGL and SK methods are consistent.

The $M_{\rm b}$-$\sigma_{\rm s}$ relation for the Osipkov-Merritt models is studied in the second panel of Fig. \ref{fig:M-sigma}. Compared to constant $\beta$ models, Osipkov-Merritt models predict higher $\sigma_{\rm s}$ for a given $M_{\rm b}$ due to radial orbits at large radii. Thus the predicted $\sigma_{\rm s}$ do not agree as well with the observed data as the constant anisotropy models. In our best fit Osipkov-Merritt models, $r_a\in [500,~1000]\kpc$, beyond which orbits become radially anisotropic for the 12 sub-samples. The observed $\sigma_{\rm s}$ suggests simple, constant, mildly radial anisotropy for the motion of satellites, rather than the  Osipkov-Merritt models with strongly radial anisotropy at large radii.

\subsection{External field effect}

In the framework of MOND, the internal acceleration is determined by both the internal baryonic distribution and the presence of a uniform external field \citep[e.g.,][]{Milgrom1983a}. That is known as the violation of the strong equivalence principle (SEP), as shown in Eq. \ref{eq:aqual}. According to this equation, internal acceleration is {reduced} by the existence of a constant external field, leading to a sharply declining rotation curve in the external field-dominated region \citep{Wu+2007}. Consequently, the phantom dark matter halo is truncated at a certain radius in systems where the external field is dominant \citep{Wu_Kroupa2015}. 
Such an external field effect (EFE) has been tested in various works, such as explaining the absence of dark matter in open clusters \citep{Milgrom1983a} and dwarf galaxies \citep[e.g.,][]{Famaey+2018,Kroupa+2018,Haghi+2019}. \citet{Chae2022ApJ} studied EFE on the RAR of disk galaxies and found EFE is stronger in AQUAL than in QUMOND. Moreover, simulations have demonstrated that the mass of phantom dark matter halo in a satellite galaxy varies as it moves along a radial orbit, due to the spatial dependence of the external field induced by the central galaxy \citep{Wu_Kroupa2013b,Wu_Kroupa2015}. As a result, in a collision, the gravitation perturbation introduced by a dwarf galaxy is weakened by the gravitational field from the target galaxy \citep{Ma_Wu2024}. The external field causes $\sigma_{\rm LoS}$ to decrease in large radii, thus it is important to study the EFE in this work. 

To include the external field, {for a spherically symmetric system}, the equation of QUMOND needs to be rewritten as follows,
\beq\label{qefe}
{\bf g} (r)+{\bf g}_{\rm ext}=\nu(\mathrm{Y}') \left[{\bf g}_{\rm N}(r)+{\bf g}_{\rm N,ext}\right],~~~~{\mathrm{with}~} \mathrm{Y}'=\frac{|{\bf g}_{\rm N}(r)+{\bf g}_{\rm N,ext}|}{a_0},
\eeq
where ${\bf g}_{\rm N,ext}$ and ${\bf g}_{\rm ext}$ are the Newtonian and Milgromian accelerations induced by the source of the external field, respectively.
{Following the methodology outlined by \cite{klypinTESTINGGRAVITYMOTION2009}, the angle $\theta$ between the external and internal acceleration vectors for a galaxy is averaged to maintain spherical symmetry
\bey \overline{{g}_{\rm N}(r)+{g}_{\rm N,ext}} &=& \int_0^\pi \sqrt{(g_{\rm N}+g_{\rm N,ext}\cos\theta)^2+(g_{\rm N,ext}\sin \theta)^2}\ {\rm d}\theta \nonumber \\ 
&=& 2\bigg|{g}_\mathrm{N}-{g}_\mathrm{N,ext}\bigg|\mathcal{E}\left[-\frac{4g_\mathrm{N}g_\mathrm{N,ext}}{(g_\mathrm{N}-g_\mathrm{N,ext})^2}\right], \label{oureqn}\eey
where $\mathcal{E}$ is the complete elliptic integral of the second kind. Thus, we can derive
\beq {\bf g}(r)=\nu(\mathrm{Y}') {\bf g}_{\rm N}(r), ~~~{\rm with} ~~~
\mathrm{Y}'= \frac{\overline{{ g}_{\rm N}(r)+{ g}_{\rm N,ext}}}{a_0}. \label{efenu}\eeq

On the other hand, for a point mass particle in an external field, BZ22 presented a semi-analytic fitting function for the internal acceleration based on numerical simulations in their Eqs. 35 and 38 (originally from \citealt{Zonoozi+2021} and reviewed by BZ22). 
We compare our analytical internal gravitational acceleration (Eqs. \ref{oureqn}-\ref{efenu}) with theirs within different strengths of $g_{\rm ext}$ in the rightmost panel of Fig. \ref{fig:M-sigma}. 
There is little difference between our results and theirs. We will show the $M_b$-$\sigma_s$ relation by considering an external field using the formulations of our Eqs. \ref{oureqn}-\ref{efenu} and those in BZ22 for a comparison as well.}

The samples selected in this work are central galaxies, and the external fields are introduced by their nearby massive galaxies or large-scale structures. Considering the typical scale of a cluster is a few $\Mpc$, the acceleration from an external field, $g_{\rm ext}$, caused by a massive nearby central galaxy with $M_{\rm b}=6\times 10^{11}\msun$ (which is the upper limit of baryonic mass in our samples of galaxies), is approximately $\sqrt{GM_{\rm b} a_0}/2\Mpc \simeq 0.01a_0$. The $g_{\rm ext}$ caused by the nearby massive objects is generally weaker than $0.01~a_0$. For instance, the Milky Way galaxy suffers an external field from M31 of this strength \citep{Wu+2008,Banik_Zhao2018}. {Besides the external field induced by the M31, the Milky Way Galaxy experiences a gravitational attraction from the large-scale structure, the so-called Great Attractor \citep{Radburn-Smith+2006}, in the direction of Sun–Galactic Centre. While precisely estimating the strength of the external field generated by the Great Attractor is challenging, an approximation can be made by the product of the Hubble constant and peculiar velocity of the Local Group, $g_{\rm ext}\approx H_0\cdot v_{\rm LG}\approx 0.01a_0$ \citep{Wu+2008,Banik+2018}, where the peculiar velocity of the Local Group is $v_{\rm LG} =627\pm 22~\kms$ \citep{Kogut+1993}. The external field introduced by the large-scale structures for the Milky Way can be higher, $g_{\rm ext} \approx 0.055a_0$ \citep{Haslbauer+2020}. This strength is constrained using the density profile of the KBC void \citep{Keenan+2013}, the local Hubble parameter, and the deceleration parameter, derived from supernovae and strong lensing systems. Since the stronger strength of external field for the Milky Way by considering the constraints of KBC void is beyond the scope of this study, we will not delve into it further.} 

We fit the ESD profiles by introducing a $g_{\rm ext}$ of $0.01a_0$ for the central galaxies and display them with dotted curves in Fig. \ref{fig:ESD_disp}, {black from our analytic formulations and red from BZ22. The ESD profiles are almost indistinguishable by using the two methods}. The presence of an external field causes the ESD profiles to drop faster in the outermost regions of galaxies in the low-mass sub-samples, including SF 1-3 and Total 1-3, opposite to the effect of the two-halo term in the $\Lambda$CDM framework. In these low-mass sub-samples, the external fields are typically much weaker than $0.01a_0$. If the external fields were stronger, it would imply the presence of a massive galaxy with a {baryonic mass of $ \frac{(g_{\rm ext}\cdot D)^2}{G a_0} \approx 10^{11}\Msun$ within a distance, $D$,} of $1\sim 2 \Mpc$ of these low-mass ``central galaxies''. This scenario contradicts the definition of central galaxies, as they would instead be identified as satellites. {But the large-scale structure might induce an external field of this strength.} In sub-sample SF 4, the ESD profile behaves the same as those in the low-mass sub-samples when $g_{\rm ext}$ is introduced. The stellar mass of this sub-sample is similar to the Milky Way. The rapidly dropping ESD profile at large radii agrees better with the observations, compared to the case of the isolated model.
On the other hand, the external field does not change the model-predicted ESD profiles in the rest of the sub-samples. 

We present the $M_{\rm b}$-$\sigma_{\rm s}$ relationship in a model incorporating an external field of $g_{\rm ext}=0.01a_0$ and an Osipkov-Merritt anisotropy (refer to the {third and fourth panels} of Fig. \ref{fig:M-sigma}), {corresponding to the internal accelerations calculated by the formulations of ours and BZ22, respectively. The $M_{\rm b}$-$\sigma_{\rm s}$ relations appear in the two panels appear rather similar.} 
Compared to the isolated Osipkov-Merritt anisotropy model (the second panel), the $\sigma_{\rm s}$ values for the low-mass sub-samples are significantly suppressed. Consequently, the fitted $M_{\rm b}$-$\sigma_{\rm s}$ relation aligns well with the observations in the low-mass sub-samples. However, the external field does not sufficiently decrease $\sigma_{\rm s}$ for a given $M_{\rm b}$ at the high-mass end. 
In addition, the $\sigma_{\rm s}$ values in the same mass range are calculated for an isotropic model combined with $g_{\rm ext}=0.01a_0$ (the solid black curve in the {third and fourth panels} of Fig. \ref{fig:M-sigma}). {These isotropic models also well reproduce the observed $M_{\rm b}$-$\sigma_{\rm s}$ relation.} 

In summary, isolated models with constant mildly radial anisotropies or isotropy predict an $M_{\rm b}$-$\sigma_{\rm s}$ relation that aligns well with observations. {When incorporating an external field, an isotropic model also agrees with observations.} 
However, introducing Osipkov-Merritt anisotropy to isolated models does not match the observed $M_{\rm b}$-$\sigma_{\rm s}$ relation. While the addition of an external field improves the fit, discrepancies remain at the high-mass end. Thus sophisticated models such as the Osipkov-Merritt anisotropy are unnecessary. The simple models with mildly radial anisotropy or isotropy in \S \ref{sk} are sufficient to explain the observed data in MOND.

\section{Conclusions}\label{conclusions}
In the framework of MOND, the baryonic mass distribution of galaxies fully determines the gravitational fields. In this study, we use two methods - galaxy-galaxy weak lensing and satellite kinematics - to estimate the baryonic mass of the galaxy sub-samples in MOND, based on observational data across a wide radial range of galaxies. The samples of galaxies are divided into two: the star-forming and the total samples. Each sample of galaxies is further segmented into six sub-samples according to their stellar mass ranges. 

Using the GGL method, we have successfully obtained the density profile parameters for cold and hot gaseous halos for each sub-sample. We have found that in all of the sub-samples of galaxies, the power law index of the hot gas density distribution is almost a constant, $\gamma=1.6$. Thus the density of the hot gas halos is close to the Plummer density distribution. This seems distinct from the X-ray observations of the hot gas halo around the Milky Way Galaxy, in which $\gamma\approx 2/3$ \citep{nicastroXRayDetectionGalaxy2023}. However, after fitting the ESD profiles with a fixed value of $\gamma\approx 2/3$, we find that it is indistinguishable from the fitted ESD profiles with $\gamma =1.6$ for the sub-samples of galaxies in similar stellar mass ranges {to} the Milky Way, due to the large uncertainties of the ESD data. 

According to the density distribution profiles, the $M_{\rm g}$ and the $r_{\rm vir}$ of the central galaxies can be computed iteratively. {The stellar fraction}, $\conveff$, in the 12 sub-samples has been derived. We have found that the values of $\conveff$ are generally higher for the star-forming galaxies than those for the total sample of galaxies at the same stellar mass. This agrees with what has been found in the $\Lambda$CDM framework \citep{Zhang+2022}. Moreover, we have calculated the rotation curves of the 12 sub-samples of galaxies. The rotation curves tend to be flat beyond a few tens of $\kpc$ for most sub-samples, except for sub-sample Total 6. 

Furthermore, we have calculated the $M_{\rm b}$ for the 12 sub-samples of galaxies using SK. Meanwhile, we have obtained the $\sigma_{\rm LoS}(r)$ profiles with constant anisotropy for each sub-sample. The baryonic masses obtained from SK agree with those from the GGL method. Combining the two methods, we have demonstrated that a $M_{\rm b}$-$\sigma_{\rm s}$ relation exists in MOND. 

Finally, more sophisticated models including an EFE and the Osipkov-Merritt anisotropy profile have been studied.  
When introducing the Osipkov-Merritt anisotropy, the $M_{\rm b}$-$\sigma_{\rm s}$ relations do not match the observations {for both isolated systems and systems embedded in an external field.} 
A simple, isolated MOND model with mildly radial anisotropy or isotropy, {or an isotropic model considering the EFE} is sufficient to explain the observed data.

The strong concordance between these two methods suggests that, within the MOND framework, GGL signals serve as reliable indicators of the dynamical mass of central galaxies. The GGL can be used to constrain the distribution of missing baryons around galaxies, and thus provide a probe to test the {behaviour of} gravity.

\section*{Acknowledgements}
We sincerely thank the anonymous reviewers for their valuable feedback and constructive suggestions. We thank Dr. HongSheng Zhao for the suggestions for the Osipkov-Merritt anisotropy models in the early stage of this work. X.W. thanks the financial support from the Natural Science Foundation of China (Number NSFC-12433002 and NSFC-12073026).  We would like to express our gratitude to Prof. Yipeng Jing's Academician Workstation for their invaluable support and contributions. We thank ``the Fundamental Research Funds for the Central Universities'' (WK3440000004). HYW is supported by CAS Project for Young Scientists 
in Basic Research (No. YSBR-062). All the authors thank Cyrus Chun Ying Tang Foundations and the 111 Project for ``Observational and Theoretical Research on Dark Matter and Dark Energy'' (B23042).



\bibliographystyle{aasjournal}
\bibliography{ref.bib}


\end{document}